\documentclass[prl,twocolumn,preprintnumbers,amsmath,amssymb,amsfonts]{revtex4}
\usepackage{color}
\usepackage{refstyle}
\usepackage{mathrsfs}
\usepackage{esint}
\usepackage{bm}
\usepackage{cancel}
\usepackage[normalem]{ulem}
\usepackage{graphicx}
\usepackage{array}
\usepackage[caption=false]{subfig}
\usepackage[unicode=true,pdfusetitle,bookmarks=true,
bookmarksnumbered=false,bookmarksopen=false,breaklinks=false,
pdfborder={0 0 0},backref=false,colorlinks=true,citecolor=red]
{hyperref}
\usepackage{cleveref}
\providecommand\eqref[1]{\ref{eq:#1}}
\renewcommand\b[1]{{\bf  #1}}
\renewcommand\vec[1]{\boldsymbol{#1}}
\renewcommand\phi{\varphi}

\newcommand\del{\nabla}
\newcommand\dd{\mathrm{d}}

\newcommand{\dbar}{{\mathrm{d}\mkern-7mu\mathchar'26\mkern-2mu}}
\allowdisplaybreaks[3]

\begin{document}
\title{Hidden entropy production and work fluctuations in an ideal active gas}
\author{Suraj Shankar$^{a,b}$}
\email{sushanka@syr.edu}
\author{M. Cristina Marchetti$^{a,b}$}
\email{mcmarche@syr.edu}
\affiliation{$^a$Physics Department and Syracuse Soft and Living Matter Program, Syracuse University, Syracuse, NY 13244, USA.\\
$^b$Kavli Institute for Theoretical Physics, University of California, Santa Barbara, CA 93106, USA.}

\date{\today}
\begin{abstract}
	Collections of self-propelled particles that move persistently by continuously consuming free energy are a paradigmatic example of active matter. In these systems, unlike Brownian ``hot colloids'', the breakdown of detailed balance yields a continuous production of entropy at steady state, even for an ideal active gas. We quantify the irreversibility for a non-interacting active particle in two dimensions by treating both conjugated and time-reversed dynamics. By starting with underdamped dynamics, we identify a hidden rate of entropy production required to maintain persistence and prevent the rapidly relaxing momenta from thermalizing, even in the limit of very large friction. Additionally, comparing two popular models of self-propulsion with identical dissipation on average, we find that the fluctuations and large deviations in work done are markedly different, providing thermodynamic insight into the varying extents to which macroscopically similar active matter systems may depart from equilibrium.
\end{abstract}
\maketitle
\textit{Introduction.} What is irreversible in active matter? These systems are driven out of equilibrium by the continuous and sustained consumption of free energy at the microscopic scale \cite{ramaswamy2010mechanics,marchetti2013hydrodynamics,ramaswamy2017active}, but quantifying such irreversibility is challenging. The persistent motion of \emph{E. coli} performing run and tumble \cite{schnitzer1993theory,berg2008coli} or of synthetic active colloids propelled by auto-phoresis \cite{paxton2006chemical,howse2007self} are classic examples of motion that breaks microscopic detailed balance by virtue of self-propulsion \cite{cates2012diffusive}, yet is diffusive on large scales. The detailed balance violations due to persistence often do not survive coarse-graining (even in the presence of weak external fields). This restores an effective equilibrium picture on large scales, thereby allowing a dilute gas of self-propelled particles to be essentially treated as a gas of  ``hot colloids'' \cite{tailleur2009sedimentation} with an effective temperature \cite{loi2008effective,palacci2010sedimentation,szamel2014self,ginot2015nonequilibrium}. In characterizing detailed balance violations on a coarse-grained scale, even manifestly non-equilibrium phenomena, such as condensation in the absence of attraction \cite{tailleur2008statistical,cates2015motility}, may then be understood by comparing them to the ``nearest'' equilibrium like model at the same scale \cite{nardini2017entropy}.

\begin{table}[t]
	\centering
	{
		{\setlength{\extrarowheight}{4pt}
		\begin{tabular}{c | c c}
			\hline
			\hline
			\hspace{0.5em} $\langle\Delta\dot{s}\rangle$ \hspace{1em} & \hspace{0.5em} Overdamped \hspace{1em} & Underdamped\hspace{0.5em}\\[4pt]
			\hline\\[-8pt]
			TRS odd propulsion\hspace{0.5em} & 0 & $\dfrac{v_0^2\gamma D_R}{T(\gamma+D_R)}$\\[8pt]
			TRS even propulsion\hspace{0.5em} & $\dfrac{v_0^2\gamma}{T}$ & $\dfrac{v_0^2\gamma^2}{T(\gamma+D_R)}$\\[8pt]
			\hline
			\hline
	\end{tabular}}
	}
	\caption{A summary of the average entropy production rate $\langle\Delta\dot{s}\rangle$ for various cases, applicable to both non-interacting ABP and AOUP (using $T_a=v_0^2\gamma/2D_R$).~The difference between the results obtained with underdamped and overdamped dynamics represents the hidden entropy production.}
	\label{table:summary}
\end{table}

	To quantify irreversibility of an ideal active gas, we examine here the microscopic dynamics of an individual active particle and evaluate the entropy production rate $\langle\Delta\dot{s}\rangle$ in two popular simple models of self-propelled particles in two dimensions ($2d$):
Active Brownian particles (ABPs) where the propulsive force has fixed magnitude and its direction is randomized by rotational noise, and active Ornstein-Uhlenbeck particles (AOUPs) where self-propulsion is modeled as a Gaussian colored noise.
Entropy production provides a direct measure of the breakdown of time-reversal symmetry (TRS) at steady state. We show below that it crucially hinges on whether the propulsive force is treated as even under TRS~\cite{speck2016stochastic,10.1088/1751-8121/aa91b9}, appropriate for active phoretic colloids, vibrated rods, or swimming bacteria, where the direction of motility encodes a physical asymmetry of the microscopic active unit, or as odd under TRS~\cite{ganguly2013stochastic,chaudhuri2014active,speck2017stochastic}, corresponding to the so-called \emph{conjugated} dynamics \cite{seifert2012stochastic}. Previous work has used both prescriptions, as well as techniques that leave the sign under TRS unspecified~\cite{fodor2016far,marconi2017heat,mandal2017entropy,puglisi2017clausius}, all with differing and sometimes conflicting notions of dissipated heat and its relation to entropy production. Additionally, a single active particle has often been found to have vanishing entropy production \cite{fodor2016far,marconi2017heat,mandal2017entropy,puglisi2017clausius,speck2017stochastic}, seemingly suggesting equilibrium behavior. We show that some of these issues can be clarified by using \emph{underdamped} dynamics along with thermal noise and taking the large friction limit only at the end, because for both TRS prescriptions the fast momenta degrees of freedom are responsible for a finite hidden entropy production \cite{celani2012anomalous,kawaguchi2013fluctuation,chun2015hidden,esposito2012stochastic}, thereby demonstrating that a single active particle is thermodynamically irreversible. This is most evident for the case of conjugated dynamics where the hidden $\langle\Delta\dot{s}\rangle$ is the only contribution, while it is subdominant at large friction for TRS even propulsive forces (see Table \ref{table:summary}). If, in contrast, inertia is neglected from the outset, a single active particle behaves as a passive colloid pulled by an external force (TRS even propulsion) or as a colloid moving at the velocity of the solvent in a sheared fluid~\cite{speck2008role,speck2017stochastic} (propulsion here is the solvent velocity, which is TRS odd), with $\langle\Delta\dot{s}\rangle=0$. This result holds for both ABP and AOUP, thereby not distinguishing the two models on the average.

We then show that the non-equilibrium nature of active particles becomes evident in the \textit{fluctuations} of thermodynamic quantities. By comparing the ABP and the AOUP models, we find that even though they have the same long-time dynamics and dissipate identically on average, their work fluctuations are vastly different. We demonstrate in a precise fashion that the AOUP gas is always further away from equilibrium compared to the ABP gas, for the same motility and persistence. Specifically, the variance of the cumulative work done to propel the particles, corresponding to the Fano factor, is strongly enhanced by activity over its linear response value for the AOUP, but not for the ABP. Our work can be extended to thermodynamic quantities of interacting active systems along with their fluctuations that are beginning to be accessible experimentally \cite{kumar2011symmetry,argun2016non,kumar2015anisotropic,battle2016broken,fodor2016nonequilibrium}.

\paragraph{The models.} We consider an underdamped active particle and set the mass and Boltzmann factor to unity. The particle velocity $\dot{\b{r}}=\b{p}$ obeys a Langevin equation,
\begin{equation}
	\dot{\b{p}}=-\gamma\b{p}+{\b{f}}_{p}+\sqrt{2T\gamma}~\vec{\xi}(t)\ ,\label{eq:abpeqn}
\end{equation}
where $\gamma$ is the friction, $T$ the temperature of the environment providing a heat bath, and $\vec{\xi}(t)$ a delta-correlated Gaussian white noise. For ABP the propulsive force $\b{f}_p=\gamma v_0\hat{\b{e}}$ has fixed magnitude, with $v_0$ the self-propulsion speed, and direction randomized by rotational noise, $\langle\hat{\b{e}}(t)\cdot\hat{\b{e}}(0)\rangle=e^{-|t|D_R}$.
	For AOUP the propulsive force is an Ornstein-Uhlenbeck process, $D_R^{-1}\dot{\b{f}}_p=-\b{f}_p+\sqrt{2\gamma T_{a}}\vec{\eta}(t)$ [$\vec{\eta}(t)$ white noise and $T_a$ an active temperature], so that $\langle\b{f}_p(t)\cdot\b{f}_p(0)\rangle=2\gamma T_aD_Re^{-|t|D_R}$. Both types of particles are diffusive at long times, with diffusivity $D=(T+T_{a})/\gamma$, where for ABP, $T_{a}=v_0^2\gamma/(2D_R)$.
	It has been shown that the large-scale phenomenology of the two models is similar even in the presence of strong interactions \cite{fily2012athermal,farage2015effective} where they both exhibit motility-induced phase separation. Yet, as we shall show below, their thermodynamic fluctuations are markedly different even at the single particle level.

\begin{figure*}[]
	\centering
	\includegraphics[width=0.9\textwidth]{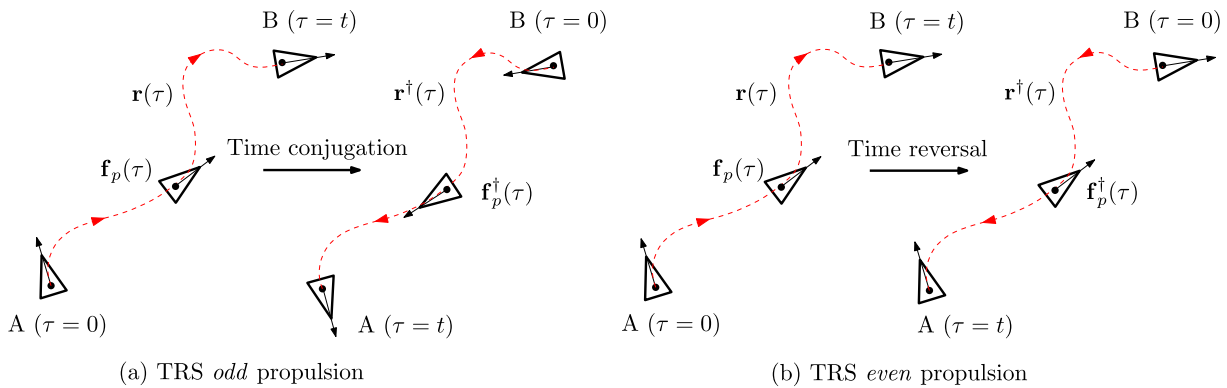}
	\caption{A cartoon of the trajectories under (a) time conjugated dynamics ($\b{f}_p$ is TRS odd) and (b) time-reversed dynamics ($\b{f}_p$ is TRS even) for a polar self-propelled particle.}
	\label{fig:TRScartoon}
\end{figure*}

\paragraph{Mean entropy production.} Irreversibility can be quantified through dissipation and entropy production, which can be calculated within the framework of stochastic thermodynamics \cite{seifert2012stochastic}. At steady state, the total entropy production of the system equals the entropy flux to the environment (also called entropy production of the medium \cite{seifert2005entropy}). For a time interval $[0,t]$, it is given by \cite{lebowitz1999gallavotti}
\begin{equation}
	\Delta s(t)=\ln\left(\dfrac{P[\b{x}(t)|\b{x}(0)]}{P^{\dagger}[\b{x}^{\dagger}(t)|\b{x}^{\dagger}(0)]}\right)\ ,\label{eq:ds}
\end{equation}
	where $\b{x}=\{\b{r},\b{p},\b{f}_p\}$ and $P[\b{x}(t)|\b{x}(0)]$ is the conditional probability of starting at $\b{x}(0)$ at time $\tau=0$ and reaching $\b{x}(t)$ at time $\tau=t$ along a given trajectory $\b{x}(\tau)$. The $\dagger$ denotes time reversal. The conditional probability for observing a forward trajectory $\b{x}(\tau)$ ($\tau\in[0,t]$) is formally written as $P[\b{x}(t)|\b{x}(0)]\propto e^{-\mathcal{A}}\prod_{\tau=0}^t\delta(\partial_{\tau}\b{r}-\b{p})$, where $\mathcal{A}[\b{x}(\tau)]$ is the Onsager Machlup functional \cite{onsager1953fluctuations} (neglecting unimportant additive constants \footnote{The dynamics of $\b{f}_p$ in both models also contributes terms to $\mathcal{A}$ but are not consequential for our present discussion. In the AOUP model, this leads to an additional rate of entropy production $\Delta\dot{s}_{R}=\b{f}_p\cdot(\b{f}_p-\sqrt{2\gamma T_a}\vec{\eta}(t))/(\gamma T_a)$, absent in the ABP model. At steady state, $\langle\Delta\dot{s}_R\rangle=0$ and it decouples from the rest of the dynamics, so we don't consider it any further.}), given by
\begin{equation}
	\mathcal{A}=\dfrac{1}{4T\gamma}\int_0^t\dd\tau\left[\partial_{\tau}\b{p}+\gamma\b{p}-\b{f}_p\right]^2\;.\label{eq:A}
\end{equation}
For non-interacting particles, the Hamiltonian of the system only involves the kinetic energy ($\mathcal{H}=\b{p}^2/2$) and the first law takes the form (in Stratanovich convention) \cite{sekimoto1998langevin}
\begin{equation}
	\dd\mathcal{H}=\b{p}\cdot\dd\b{p}=\dbar w-\dbar q\;,\label{eq:firstlaw}
\end{equation}
where $\dbar w$ is the propulsive work done and $\dbar q$ is the heat dissipated into the reservoir. The sign convention used is that both heat dissipated into the bath and work done by the environment on the system are taken to be positive. Requiring the Clausius relation, we equate $\dbar q(t)=T\Delta s(t)$, which as we will see below is consistent with Sekimoto's \cite{sekimoto1998langevin} definition of heat only for the TRS even case. It is clear from Eq.~\ref{eq:ds} that, as discussed in the Introduction, entropy production depends on whether the propulsion is treated as a force (hence TRS even) or as a velocity (hence TRS odd). We discuss both cases here, although the TRS even prescription is more directly relevant to physical realizations. Also, the calculation of the mean entropy production is outlined here for ABP. The result turns out to be the same for AOUP.

\underline{TRS odd propulsion.} The prescription of conjugated dynamics ($\b{r}^{\dagger}(\tau)=\b{r}(t-\tau)$, $\b{p}^{\dagger}(\tau)=-\b{p}(t-\tau)$ and $\b{f}_p^{\dagger}(\tau)=-\b{f}_p(t-\tau)$ on a time interval $\tau\in[0,t]$, see Fig.~\ref{fig:TRScartoon}(a)) most clearly illustrates the importance of retaining the fast momenta degrees of freedom and the associated hidden entropy production.~Considering from the outset overdamped dynamics and treating motility as a TRS odd velocity seems to lead identically to $\Delta\dot{s}=0$, in the absence of interactions \cite{fodor2016far,speck2017stochastic}, wrongly suggesting that the system is in equilibrium~\footnote{Note that the procedure of Refs.~\cite{fodor2016far,marconi2017heat,mandal2017entropy} cannot be used in the presence of translational noise with $T\neq 0$.}.
~Working instead with the underdamped equations, we obtain the entropy production rate to be $\Delta\dot{s}=-\dot{\b{p}}\cdot(\b{p}-v_0\hat{\b{e}})/T$. Averaging over noise, in steady state, we get
\begin{equation}
	\langle\Delta\dot{s}\rangle=\dfrac{v_0^2\gamma D_R}{T(\gamma+D_R)}=\dfrac{v_0^2}{T}D_R+\mathcal{O}\left(\dfrac{D_R}{\gamma}\right)\ .
\end{equation}
This demonstrates a hidden entropy production in active matter arising from the entropic cost to maintain a finite persistence and evade thermalization of the fast momentum. By taking the overdamped limit at the very outset, i.e., $t\gg\gamma^{-1}$, the momentum is implicitly assumed to have relaxed to the equilibrium Maxwell-Boltzmann distribution, but this is simply not true on time scales of $\mathcal{O}(D_R^{-1})$ due to the persistence of motion. As the momentum of the active particle is effectively slaved to the motility, on short time scales ($\sim\gamma^{-1}$) it relaxes to the stationary non-equilibrium distribution $P_{ss}(\b{p}|\hat{\b{e}})\propto\exp(-|\b{p}-v_0\hat{\b{e}}|^2/2T)$  \cite{baskaran2010nonequilibrium}. On time scales $\sim D_R^{-1}(>\gamma^{-1})$, the polarization direction decorrelates, but it also forces the momentum to do the same in tandem, an act that requires work to be done and dissipated irreversibly. For $\gamma/D_R\gg 1$, one can also view $\langle\Delta\dot{s}\rangle$ as the symmetrized relative entropy (or the symmetrized Kullback-Leibler divergence \cite{kullback1951information}),
\begin{equation}
	\Delta s_{\mathrm{rel}}=-\int\dfrac{\dd\hat{\b{e}}}{2\pi}\int\dd^2p\left[P_{\mathrm{eq}}(\b{p})-P_{ss}(\b{p}|\hat{\b{e}})\right]\ln\left(\dfrac{P_{ss}(\b{p}|\hat{\b{e}})}{P_{\mathrm{eq}}(\b{p})}\right)\ ,
\end{equation}
dissipated to the bath in a rotational correlation time $D_R^{-1}$, with $P_{\mathrm{eq}}(\b{p})\propto\exp(-p^2/2T)$. For $D_R=0$, the system behaves as if it were in a background steady deterministic flow and $\langle\Delta\dot{s}\rangle$  vanishes.

\underline{TRS even propulsion.} If motility is treated as a TRS even \emph{non-conservative force} (Fig.~\ref{fig:TRScartoon}(b)), a single active particle is then analogous to a driven colloid. In this case $\b{r}$ and $\b{p}$ transform as before under time reversal, but $\b{f}_p^{\dagger}(\tau)=\b{f}_p(t-\tau)$. Using Eqs.~\ref{eq:A} and~\ref{eq:ds}, the entropy production rate is identified as $\Delta\dot{s}=\b{p}\cdot(\gamma\b{p}-\sqrt{2T\gamma}\vec{\xi})/T$. The rate of heat dissipated $\dot{q}=T\Delta\dot{s}$ is as expected with $\b{p}=\dot{\b{r}}$ \cite{sekimoto1998langevin} and the rate of work done (from Eq.~\ref{eq:firstlaw}) is given by $\dot{w}=v_0\gamma\hat{\b{e}}\cdot\b{p}$, which is the power injected by the propulsive force $\b{f}_p$. At steady state, the average rate of dissipation is
\begin{equation}
	\langle\dot{q}\rangle=\langle\dot{w}\rangle=\dfrac{v_0^2\gamma^2}{\gamma+D_R}\simeq v_0^2\gamma\left[1+\mathcal{O}\left(\frac{D_R}{\gamma}\right)\right]\;.\label{eq:qw}
\end{equation}
For $\gamma\gg D_R$ the mean dissipation rate is the same as for a particle dragged by a constant force $v_0\gamma$. Starting from the outset with overdamped equations yields identically $\langle\dot{q}\rangle=\langle\dot{w}\rangle=v_0^2\gamma$. Therefore when self-propulsion is treated as a TRS-even force all hidden entropy contributions only appear at sub-leading order in $D_R/\gamma$.

The mean entropy production rate for the various combinations considered here is summarized in Table~\ref{table:summary} \footnote{These results can easily be extended to anisotropic friction $\vec{\gamma}=\gamma_{||}\hat{\b{e}}\hat{\b{e}}+\gamma_{\perp}(\b{1}-\hat{\b{e}}\hat{\b{e}})$, where for example, the average dissipation rate for a TRS even propulsion is $\langle\dot{q}\rangle=v_0^2\gamma^2_{||}/(\gamma_{||}+D_R)$.}. Identifying $T_a=v_0^2\gamma/2D_R$ relates the AOUP model to the ABP, highlighting that both models have the same mean dissipation rate at steady state. So the two models are thermodynamically identical \emph{on average}.

\paragraph{Work fluctuations.} The difference between the two models and true non-equilibrium nature becomes apparent in their \emph{fluctuations}. We compute the variance of the cumulative work $\Delta w(t)=\int_0^t\dd\tau~\dot{w}(\tau)$ done in propelling the active particle for a time $t$. Here, we consider only the physically relevant TRS-even case. At long times ($t\to\infty$), we have
\begin{equation}
	\langle\Delta w(t)^2\rangle-\langle\Delta w(t)\rangle^2=2T_w\langle\Delta w(t)\rangle\ ,\label{eq:Tw}
\end{equation}
where $T_w$ (the Fano factor) is an effective temperature for work fluctuations (distinct from the active temperature $T_a$). One can compute $T_w$ through a Green-Kubo like formula, relating it to the time auto-correlation of the power input,
\begin{align}
	T_w=\dfrac{1}{\langle\dot{w}\rangle}\int_0^{\infty}\dd t\left[\langle\dot{w}(t)\dot{w}(0)\rangle-\langle\dot{w}\rangle^2\right]\ .
\end{align}
As $T_w$ quantifies the relative fluctuations of $\dot{w}$, a current, it obeys a universal bound at steady-state, $T_w\geq T$, first conjectured for out-of equilibrium reaction networks \cite{barato2015thermodynamic} and later proven in a general stronger form by \citet{gingrich2016dissipation}. A remarkable result, the universal bound provides an uncertainty relation between current fluctuations and dissipation, generalizing equilibrium fluctuation-dissipation theorems \cite{kubo1966fluctuation} to far from equilibrium steady states.

\begin{figure}[]
	\centering
	\includegraphics[width=0.5\textwidth]{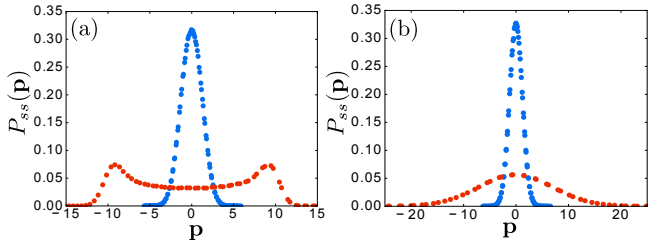}
	\caption{The steady state probability distribution of the particle momentum is plotted for (a) the ABP model with $v_0=1$ (blue) and $v_0=10$ (red), and (b) the AOUP model with $v_0=1$ (blue) and $v_0=10$ (red). As both $p_x$ and $p_y$ are identically distributed, they are plotted with the same color and symbol. Parameters $\gamma=100$, $D_R=1$, and $T=1$ are chosen common.}
	\label{fig:pdist}
\end{figure}

\begin{figure}[]
	\centering
	\includegraphics[width=0.52\textwidth]{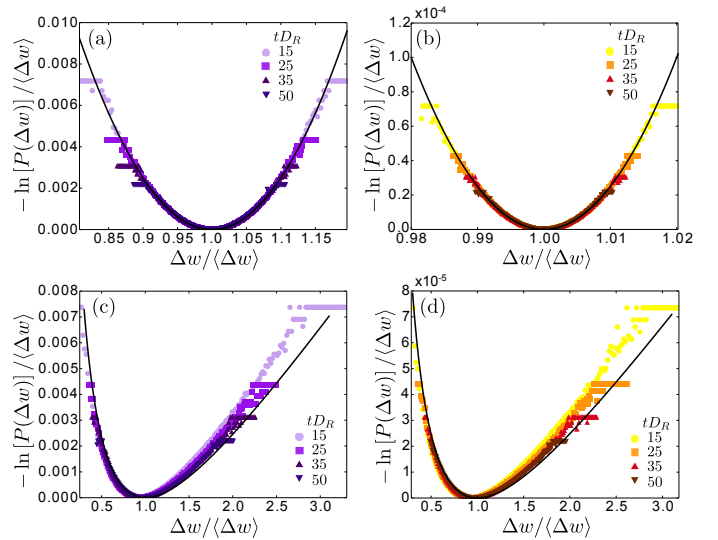}
	\caption{The large deviation function of work done in the ABP [(a) $v_0=1$, (b) $v_0=10$] and AOUP [(c) $v_0=1$, (d) $v_0=10$] models. The black lines in all four plots are the theoretical predictions from Eq.~\ref{eq:abp_ldf} and Eq.~\ref{eq:F_aoup} for the two models. The other parameters are $\gamma=100$, $D_R=1$ and $T=1$.}
	\label{fig:ldf_panel}
\end{figure}

For the underdamped ABP we find
\begin{align}
	T_w^{\mathrm{ABP}}=T+\dfrac{\langle\dot{w}\rangle D_R^2}{\gamma(\gamma+D_R)(\gamma+2D_R)}\simeq T\ ,
\end{align}
where the second equality holds for negligible inertia ($\gamma/D_R\to\infty$), i.e., the ABP \emph{saturates} the universal dissipation bound ($T_w=T$) for \emph{arbitrary} motility and persistence. An important and surprising consequence of this result is that a free overdamped ABP gas is \emph{always} within the \emph{linear} response regime from a steady state with detailed balance, regardless of what $v_0$ or $D_R$ are. This is especially counterintuitive given that for large $v_0$ the velocity distribution is non-Maxwellian and bimodal (Fig.~\ref{fig:pdist}a). Since the particle is linearly close to equilibrium, all higher cumulants of the work done vanish and one can easily compute the large deviation functional for the work current $J_t$, at steady state for large friction, with the result (see Fig.~\ref{fig:ldf_panel}a-b)
\begin{equation}
	\lim_{t\to\infty}-\dfrac{1}{t}\ln P\left(\dfrac{\Delta w(t)}{t}=J_t\right)=\dfrac{(J_t-\langle\dot{w}\rangle)^2}{4T\langle\dot{w}\rangle}\;.\label{eq:abp_ldf}
\end{equation}
In other words the  work distribution is Gaussian and satisfies a fluctuation theorem $\langle e^{-\Delta w/T}\rangle=1$ \cite{seifert2005entropy,seifert2012stochastic}
In Ref.~\cite{PhysRevLett.119.140604}, it was shown that overdamped $2d$ chiral active Brownian particles also similarly saturate the dissipation bound and are hence linearly close to equilibrium as well.

Doing the same, we compute the work fluctuations for the AOUP, with the result
\begin{equation}
	T_w^{\mathrm{AOUP}}=T+T_a+\dfrac{\langle\dot{w}\rangle}{2(D_R+\gamma)}\ .
\end{equation}
Unlike the ABP, the AOUP model \emph{does not} saturate the universal bound on dissipation in the limit of large friction. In fact, $T_w^{\mathrm{AOUP}}\simeq T+T_a$ (for $\gamma\gg D_R$) \footnote{Here we take the large friction limit at fixed $T_a$. Keeping a putative self-propulsion speed $v_0=\sqrt{2T_aD_R/\gamma}$ fixed instead only results in a higher $T_w^{\mathrm{AOUP}}$.}, indicating that the system moves further way from the equilibrium steady state (and the linear response regime) with increasing active temperature $T_a$. These enhanced work fluctuations arise from the fact that the fluctuations of the propulsive force $\b{f}_p$ are unbounded for AOUP and lead to the power input being correlated on longer time-scales $\sim D_R^{-1}$ (instead of $(\gamma+D_R)^{-1}$ as for the ABP model).
Our results suggest that tracers in an active bath that are usually thought to be well described as AOUP \cite{maggi2014generalized} may be thermodynamically distinct from actual active particles.

One can also compute the large-deviation function of work done, for the AOUP model (see Ref.~\cite{SI} for the derivation). We compute the cumulant generating function $\mathcal{F}(\lambda)=-\ln\langle e^{-\lambda\Delta w(t)}\rangle/t$ as an eigenvalue of a tilted Fokker-Planck operator \cite{lebowitz1999gallavotti} using a Gaussian ansatz for the corresponding eigenfunction, with the result
\begin{equation}
	\dfrac{\mathcal{F}(\lambda)}{\gamma}=-1-\dfrac{D_R}{\gamma}+\sqrt{1+\dfrac{D_R^2}{\gamma^2}+2\dfrac{D_R}{\gamma}\sqrt{1+4T_a\lambda(1-T\lambda)}}\ .\label{eq:F_aoup}
\end{equation}
This function has branch cuts outside the interval $[\lambda_-,\lambda_+]$, with $\lambda_{\pm}=[1\pm\sqrt{1+{T}/{T_a}}]/(2T)$ leading to exponential non-Gaussian tails in the work distribution. The large-deviation function is then obtained by a Legendre transform of $\mathcal{F}(\lambda)$ and is shown in Fig.~\ref{fig:ldf_panel}c-d. A Gallavotti-Cohen like symmetry \cite{lebowitz1999gallavotti} is realized here as $\mathcal{F}(\lambda)=\mathcal{F}(T^{-1}-\lambda)$ and leads to a corresponding detailed fluctuation theorem for $P(\Delta w)$. Extreme rare fluctuations in the AOUP model are far in excess than in the ABP. As recent experiments have measured both Gaussian and non-Gaussian large deviations in a self-propelled particle \cite{kumar2011symmetry}, we expect our results can advise the thermodynamically appropriate modeling of such particles.
 It would be interesting to see how these fluctuations change when interactions are added in both models and how these results will play out when extended to coarse-grained scales. Some recent works \cite{cagnetta2017large,whitelam2018phase} have correlated large deviations in work to clustering and phase separation in interacting active systems. Even from our single particle treatment, we see that large fluctuations are controlled by the statistics of persistence (that can be modified by interactions) and encodes the time correlation of the power input $\langle\dot{w}(t)\dot{w}(0)\rangle$. A comparison including the interaction time scale in the power auto-correlation is left for future work.

\paragraph{Conclusions.} To conclude, we have argued the importance of including fast degrees of freedom in thermodynamic treatments of active matter and shown how one may gain different notions of irreversibility from conjugated and time reversed dynamics. The presence of hidden entropy production extends to other situations as well, for example, in chiral active rotors \cite{lenz2003membranes,van2016spatiotemporal} one would have to retain the fast angular-momentum as well. Additionally, in cases where self-propulsion ultimately comes from an underlying microscopic chemical reaction, the chemical variable must be retained to obtain the physical dissipation experimentally measurable in the system.
By working within a Langevin framework as in Ref.~\cite{ramaswamy2017active} we correctly reproduce \cite{SI} the recent results of \citet{10.1088/1751-8121/aa91b9}, without having to introduce a discrete lattice model. The claimed failure of the time-reversal procedure at the level of stochastic trajectories \cite{10.1088/1751-8121/aa91b9} is then seen to be a consequence of the hidden entropy production.
Finally, we emphasize the importance of going beyond average quantities and look at fluctuations of the work done in propelling two model active systems. Comparing the ABP and the AOUP models, we find that even though they have the same long-time dynamics and dissipate identically on average, their work fluctuations are vastly different signaling their distinct nonequilibrium features.



We thank Sriram Ramaswamy for helpful discussions and Andrea Puglisi for useful comments. This work was primarily supported by NSF-DMR-1609208. Additional support was provided by NSF-DGE-1068780 (MCM) and by NSF-PHY-1748958 (SS, MCM). The authors also acknowledge support of the Syracuse University Soft and Living Matter Program and thank the KITP for hospitality during part of this project.



\pagebreak
\widetext
\begin{center}
\textbf{\large Hidden entropy production and work fluctuations in an active gas\\~\\
SUPPLEMENTARY INFORMATION\\}
	\vspace{1.3em}
Suraj Shankar$^{a,b}$ and M. Cristina Marchetti$^{a,b}$\\
	\textit{$^a$Physics Department and Syracuse Soft and Living Matter Program,\\ Syracuse University, Syracuse, NY 13244, USA.\\
$^b$Kavli Institute for Theoretical Physics, University of California, Santa Barbara, CA 93106, USA.}
\end{center}


\setcounter{equation}{0}
\setcounter{figure}{0}
\setcounter{table}{0}
\setcounter{page}{1}
\makeatletter
\renewcommand{\theequation}{S\arabic{equation}}
\renewcommand{\thefigure}{S\arabic{figure}}
\renewcommand{\bibnumfmt}[1]{[S#1]}
\renewcommand{\citenumfont}[1]{S#1}
\thispagestyle{empty}

\section{I.\hspace{1em} Large deviation functional for an AOUP}
The probability distribution of work current is given by
\begin{equation}
	P\left(\dfrac{\Delta w(t)}{t}=J_t\right)=\langle\delta\left(\Delta w(t)-t J_t\right)\rangle=\int\dd\lambda\ e^{-t(\mathcal{F}(\lambda)-\lambda J_t)}\ ,
\end{equation}
where the integral is over a contour $(c-i\infty,c+i\infty)$ in the complex $\lambda$ plane where $c$ is some constant chosen such that the integral converges. The cumulant generating function $\mathcal{F}(\lambda)=-\ln\langle e^{-\lambda\Delta w(t)}\rangle/t$ is defined for all $\lambda\in\mathbb{C}$ by analytic continuation. As $t\to\infty$, the integral over $\lambda$ is dominated by the saddle point $\mathcal{F}'(\lambda_*)=J_t$. Using this, in the large time limit, we then obtain the large deviation function for the current $J_t$ to be
\begin{equation}
	\lim_{t\to\infty}-\dfrac{1}{t}\ln P\left(\dfrac{\Delta w(t)}{t}=J_t\right)=\mathcal{F}(\lambda_{*})-\lambda_*J_t\ ,
\end{equation}
with $\lambda_{*}$ inverted as a function of $J_t$.

Starting with the joint probability distribution $P(\b{p},\b{u},\Delta w;t)$, we can perform a bilateral Laplace transform only on $\Delta w$ to get
\begin{equation}
	\Psi_{\lambda}(\b{p},\b{f}_p;t)=\int\dd\Delta w\ e^{-\lambda\Delta w} P(\b{p},\b{f}_p,\Delta w;t)\ .
\end{equation}
Using the fact that $\dot{w}=\partial_t\Delta w=\b{p}\cdot\b{f}_p$, along with the AOUP equation of motion, the joint probability distribution $P(\b{p},\b{u},\Delta w;t)$ satisfies the following Fokker-Planck equation
\begin{equation}
	\partial_tP=-\b{f}_p\cdot\del_{\b{p}}P+\gamma\del_{\b{p}}\cdot\left[\b{p}P+T\del_{\b{p}}P\right]+D_R\del_{\b{f}_p}\cdot\left[\b{f}_pP+\gamma T_a D_R\del_{\b{f}_p}P\right]-\b{p}\cdot\b{f}_p\del_{\Delta w}P\ .\label{eq:fp}
\end{equation}
Laplace transforming Eq.~\ref{eq:fp}, we get the following tilted equation for $\Psi_{\lambda}(\b{p},\b{f}_p;t)$
\begin{equation}
	\partial_t\Psi_{\lambda}=-\b{f}_p\cdot\del_{\b{p}}\Psi_{\lambda}+\gamma\del_{\b{p}}\cdot\left[\b{p}\Psi_{\lambda}+T\del_{\b{p}}\Psi_{\lambda}\right]+D_R\del_{\b{f}_p}\cdot\left[\b{f}_p\Psi_{\lambda}+\gamma T_a D_R\del_{\b{f}_p}\Psi_{\lambda}\right]-\lambda\b{p}\cdot\b{f}_p\Psi_{\lambda}\ .
\end{equation}
Using a Gaussian ansatz for $\Psi_{\lambda}$ we write
\begin{equation}
	\Psi_{\lambda}(\b{p},\b{f}_p;t)\propto\exp\left(\mu_{\lambda}t-\dfrac{1}{2\sigma}\left|\b{p}-\alpha\b{f}_p\right|^2-\dfrac{\Sigma}{2}|\b{f}_p|^2\right)
\end{equation}
with $\alpha$, $\sigma$ and $\Sigma$ as undetermined constants along with the eigenvalue $\mu_{\lambda}$, which satisfy the following algebraic equations,
\begin{gather}
	\sigma=T+D_R^2T_a\alpha^2\\
	\mu_{\lambda}+2D_R^2\gamma T_a\Sigma=2D_R\\
	D_RT\Sigma(D_RT_a\gamma\Sigma-1)+\alpha^2(1+D_R^2T_a\Sigma)[\gamma+D_R(D_RT_a\gamma\Sigma-1)]=\alpha\\
	1-\lambda(T+D_R^2T_a\alpha^2)=\alpha[\gamma+D_R(2D_RT_a\gamma\Sigma-1)]
\end{gather}
Solving these equations for the largest eigenvalue $\mu_\lambda$ and demanding that $\mu_{\lambda=0}=0$ (required by normalization of the distribution function) we obtain a single consistent root,
\begin{equation}
	\mu_{\lambda}=D_R+\gamma-\sqrt{D_R^2+\gamma^2+2\gamma D_R\sqrt{1+4T_a\lambda(1-T\lambda)}}\ .
\end{equation}
At long times ($t\to\infty$), integrating (marginalizing) over $\{\b{p},\b{f}_p\}$, we have $\langle e^{-\lambda\Delta w(t)}\rangle=\int\dd\b{p}\int\dd\b{f}_p\Psi_{\lambda}\sim e^{\mu_{\lambda}t}$. So the cumulant generating function $\mathcal{F}(\lambda)=-\ln\langle e^{-\lambda\Delta w(t)}\rangle/t=-\mu_{\lambda}$ for $t\to\infty$, as is quoted in the main text. The corresponding large deviation function is then obtained by numerically inverting the Legendre transform of $\mathcal{F}(\lambda)$.

\section{II.\hspace{1em} Hidden entropy production due to a chemical reaction}

Here we generalize the calculations described in the main text by considering an isolated active particle whose self-propulsion is driven by an internal chemical mechanism. Following Ref.~\cite{SIramaswamy2017active}, we couple the particle dynamics to a pair of fast variables ($X$,$\pi$) that account for the chemical reaction coordinate and the chemical velocity, within a Langevin framework.
\begin{gather}
	\dot{\b{p}}=-\gamma\b{p}+\Gamma_{12}\hat{\b{e}}\ \partial_{\pi}\mathcal{H}+\vec{\xi}'(t)\ ,\label{eq:peqn}\\
	\dot{\pi}=-\Gamma\partial_{\pi}\mathcal{H}+\Gamma_{12}\hat{\b{e}}\cdot\b{p}-\partial_X\mathcal{H}+\nu(t)\ ,\label{eq:pieqn}
\end{gather}
where $\dot{\b{r}}=\b{p}$ is the particle velocity ($m=1$) and $\dot{X}=\partial_\pi\mathcal{H}$ is the chemical velocity. $\gamma$ and $\Gamma$ are friction terms and the polarization (or direction of motility) $\hat{\b{e}}$ is introduced here along with $\Gamma_{12}$ as a dissipative Onsager cross-coupling between the physical and chemical momenta. The dynamics of $\hat{\b{e}}$ itself is pure rotational diffusion with a persistence time $D_R^{-1}$ just as in the main text. The corresponding zero mean Gaussian white noise $\vec{\xi}'$ and $\nu$ are chosen to have correlations that respect the fluctuation-dissipation theorem. Hence, as it stands, this system of Langevin dynamics describes \emph{passive} dynamics of a single particle that relaxes to an \emph{equilibrium} state described by the Boltzmann distribution with  energy $\mathcal{H}$.

If, instead of allowing the system to relax to equilibrium, we favor the forward reaction over the reverse one by holding $\partial_{X}\mathcal{H}=-\Delta\mu$ at a fixed chemical potential difference (between reactants and products), then the equations describe an out-of-equilbrium  \emph{active} particle. Eliminating the chemical velocity $\dot{X}$ and the chemical reaction coordinate in favor of $\b{p}$, in the overdamped approximation ($\dot{\pi}\approx0$), we have
\begin{equation}
	\dot{X}=\partial_{\pi}\mathcal{H}\simeq\dfrac{\Gamma_{12}}{\Gamma}\hat{\b{e}}\cdot\b{p}+\dfrac{\Delta\mu}{\Gamma}+\dfrac{\nu(t)}{\Gamma}\ .\label{eq:Xdot}
\end{equation}
Substituting Eq.~\ref{eq:Xdot} into Eq.~\ref{eq:peqn}, we get rid of the chemical reaction and obtain an underdamped active Brownian particle with anisotropic friction,
\begin{equation}
	\dot{\b{p}}=-\vec{\gamma}\cdot\left(\b{p}-v_0\hat{\b{e}}\right)+\vec{\xi}(t)\ ,\label{eq:pwoX}
\end{equation}
where $\vec{\gamma}=\gamma_{||}\hat{\b{e}}\hat{\b{e}}+\gamma_{\perp}(\b{1}-\hat{\b{e}}\hat{\b{e}})$ with $\gamma_{||}=\gamma-(\Gamma_{12}^2/\Gamma)$ and $\gamma_{\perp}=\gamma$. The Gaussian noise $\vec{\xi}(t)$ has correlations $\langle\vec{\xi}(t)\vec{\xi}(t')\rangle=2T\vec{\gamma}\delta(t-t')$ satisfying the fluctuation-dissipation theorem. The active drive enters as self-propulsion for $\Delta\mu\neq 0$ with $v_0=\Delta\mu\Gamma_{12}/(\Gamma\gamma_{||})$.

Treating the polarization $\hat{\b{e}}$ as a physical TRS even vector, the rate of entropy production is $\Delta\dot{s}=\b{p}\cdot(\vec{\gamma}\cdot\b{p}-\vec{\xi})/T$ and the work done is $\dot{w}=v_0\gamma_{||}\hat{\b{e}}\cdot\b{p}$ -- a direct generalization of the expressions quoted in the main text. At steady-state, this leads to
\begin{equation}
	\langle\Delta\dot{s}\rangle=\dfrac{v_0^2\gamma_{||}^2}{T(\gamma_{||}+D_R)}\ .\label{eq:ds1}
\end{equation}
Instead of starting with Eq.~\ref{eq:pwoX}, where the \emph{fast} chemical variable $X$ has been eliminated, one must retain its dynamics in order to correctly capture the dissipation in the system as the chemical reaction \emph{fails to equilibriate} for $\Delta\mu\neq 0$. Just as was the case for the rapidly relaxing momenta whose non-equilibriation lead to a hidden entropy contribution, the internal chemical reaction also generates hidden entropy even at steady state. This contribution is easily obtained even from considerations of linear irreversible thermodynamics and is given by $\langle\Delta \dot{s}_{X}\rangle=\Delta\mu^2/T\Gamma$ (for large $\Gamma$). Including this hidden entropy contribution, we get the total rate of entropy production to be ($\Delta\dot{s}_{\mathrm{tot}}=\Delta\dot{s}+\Delta\dot{s}_{X}$)
\begin{equation}
	\langle\Delta\dot{s}_{\mathrm{tot}}\rangle=\dfrac{v_0^2\gamma_{||}\gamma_{\perp}}{T(\gamma_{\perp}-\gamma_{||})}\ ,\label{eq:ds2}
\end{equation}
where we have used the fact that $\gamma_{||},\gamma_{\perp}\gg D_R$ along with the relation $\gamma_{\perp}-\gamma_{||}=\Gamma_{12}^2/\Gamma$. This expression for the rate of entropy production coincides with that obtained by \citet{SI10.1088/1751-8121/aa91b9} who work with a discrete lattice model of an active particle. Eliminating the chemical reaction at the very outset leads to a reduced entropy production rate as given in Eq.~\ref{eq:ds1} which was claimed in Ref.~\cite{SI10.1088/1751-8121/aa91b9} to point to a failure of the procedure of time reversal at the level of Langevin equations. Here, from the analysis above, it is clear that this is not really true and the reason the Langevin dynamics (without the explicit reaction) gave the wrong answer is because of a hidden contribution to entropy production that arises from the non-equilibriation of the fast chemical velocity.

\end{document}